\documentclass[aps,prb,reprint,showkeys]{revtex4-2}

\usepackage{amssymb}
\usepackage{amsmath}
\usepackage{graphicx}
\usepackage{dcolumn}
\usepackage{bm}

\usepackage{xcolor}
\usepackage{csquotes}
\usepackage[version=4]{mhchem}
\usepackage{siunitx}

\usepackage{natbib}

\begin{document}

\title{The bridge function as a functional of the radial distribution function: Operator learning and application}

\date{\today}



\author{Martin Panholzer}
\thanks{Contact author: panholzer@unisoftwareplus.com}
\affiliation{Uni Software Plus GmbH, Linzer Strasse 6, 4320 Perg, Austria}

\author{Michael Haring}
\author{Thomas Wallek}
\thanks{Contact author: thomas.wallek@tugraz.at}
\affiliation{Institute of Chemical Engineering \& Environmental Technology, Graz University of Technology, Inffeldgasse 25/C, 8010 Graz, Austria}

\author{Robert E. Zillich}
\thanks{Contact author: robert.zillich@jku.at}
\affiliation{Institute for Theoretical Physics, Johannes Kepler University Linz, Altenberger Straße 69, 4040 Linz, Austria}

\begin{abstract}
Properties of classical molecular systems can be calculated with integral equation theories based on the Ornstein-Zernike (OZ) equation and a complementing closure relation. One such closure relation is the hyper netted chain (HNC) approximation, which neglects the so-called bridge function. 
We present a new way to use machine learning to train a deep operator network to predict the bridge function, based on the radial distribution function as input.
Bridge functions for the Lennard-Jones fluid are calculated from 
Monte Carlo simulations in a wide range of densities and temperatures. These results are used to train the deep operator network. 
This network is employed to improve the HNC closure by the prediction for the bridge function, and the resulting set of equations is solved iteratively.
For assessment, we compare the radial distribution function and the pressure, calculated by the viral expression, with Monte Carlo results and standard HNC.
We demonstrate that incorporating the neural network based bridge function in the closure relation leads to substantially improved predictions. 
Universality of our method is demonstrated by comparing results for the hard sphere fluid, calculated with our model trained on the Lennard-Jones fluid, with exact hard sphere results, showing overall good agreement.
\end{abstract}


\maketitle

\section{Introduction}

Understanding the structural and thermodynamic properties of atomic and molecular liquids remains a fundamental task in statistical mechanics. The integral equation theory provides an efficient means of describing these properties, with the Ornstein-Zernike (OZ) equation serving as the link between the total correlation function $h(r)$ to the direct correlation function $c(r)$ \cite{Hansen_2013}. Based on the OZ equation, various successful approaches such as the molecular Ornstein-Zernike equation (MOZ) and the Reference Interaction Site Model (RISM) have been developed and used, e.g.\ to study solvation thermodynamics \cite{Ratkova_2015}.

To solve the OZ equation, various closure relations have been proposed, among which the family of hypernetted-Chain (HNC) approximations plays a significant role \cite{Morita_1960}. HNC approximations incorporate essential correlation effects while maintaining a relatively simple mathematical structure. However, they all neglect certain higher-order many-body contributions, which are crucial for accurately modeling dense and strongly interacting fluids in a more general way. These missing contributions are captured by the bridge function, defined below, which introduces a more refined treatment of particle correlations \cite{Rosenfeld_1979}.
Diagrammatic expansions provide a systematic approach for deriving these closure relations. By considering the Mayer cluster expansion and its graphical representation, one can identify different classes of diagrams that contribute to the correlation functions. The HNC approximation emerges naturally from a subset of these diagrams, where bridge diagrams account for additional many-body effects \cite{Wertheim_1963}. Only in the limit of summing up the infinite set of all bridge diagrams, the resulting bridge function and thus integral equation theory would become exact.

The bridge diagrams are a well defined set of cluster diagrams where the connections (bonds) are the Mayer function $f(r)=e^{-\beta v(r)}-1$, where $\beta=\frac{1}{k_\text{B} T}$ and we assume a pair-wise interaction between point particles (such as interaction sites in RISM) given by $\sum_{i<j} v(r_{ij})$. Resummation of sub-diagrams can greatly reduce the number of diagrams. For the bridge diagrams this leads to the redefinition of the bonds as total correlation function $h(r)$. In this way we see that the bridge function is a functional of $h(r)$ or, equivalently, of the radial distribution function (RDF) $g(r)=h(r)+1$. Exact knowledge of the bridge functional $B(r)=B[g;r]$ would allow the exact calculation of $g(r)$ for a given interaction $v(r)$ -- and from that essentially all thermodynamic properties -- by self-consistently solving the HNCB equation (exact closure)
\begin{equation}
    g(r) = \exp\left[-\beta v(r) + \gamma(r) + B(r)\right]
    \label{eq:fullHNC}
\end{equation}
and the OZ equation
\begin{equation}
    \gamma(r) = \rho\int d^3r'\, (g(r'-r)-1) c(r) \ ,
    \label{eq:OZ}
\end{equation}
where $\gamma(r)=g(r)-1-c(r)$ is the indirect correlation function. Hence, in principle the integral equation method provides a framework for calculating thermodynamic properties of fluids exactly. The catch is that the exact bridge functional $B[g;r]$ cannot be calculated because it is an infinite sum of Mayer cluster diagrams. Usually, $B[g;r]$ has been approximated, leading to a host of approximate closures of the OZ equation \cite{percusPR58,verletMolPhys81,martynovMolPhys83,rogersPRA84,zerahJCP86,duhJCP95,kovalenkoJCP99,martynovJCP99}. Unfortunately, these approximations are usually only suitable for a certain type of liquid (Coulomb \cite{Baus_1980}, hard spheres (HS), Lennard-Jones (LJ), etc.). The trivial approximation is to omit $B$ completely, $B(r)=0$, termed HNC0 approximation.

Monte Carlo (MC) and Molecular Dynamics (MD) simulations serve as approaches complementary to integral equation theory by providing numerical benchmarks for correlation functions and thermodynamic properties \cite{Frenkel_2023}, but require an order of magnitude higher computational effort than integral equation methods. Molecular simulations allow for direct sampling of particle configurations and can be used to validate theoretical predictions obtained integral equation theories.

Recently, various attempts have been made to link integral equation theories and molecular simulations with the machine learning (ML) methods from the rapidly evolving field of artificial intelligence (AI). Such methods have been proposed for various goals, such as to improve the results of established RISM closures with ML-based methods, to directly model RDFs and correlation functions, to model bridge functions, or to extract force-field parameters as part of the so-called \enquote{inverse problem} of statistical mechanics \cite{Goodall_2021,Wu_2023}.
\paragraph*{ML-based improvement of RISM results.} Osaki et al.\ proposed the 3D-RISM-AI method, using the hydration free energy based on the 3D-RISM method as input feature for an ML approach to predict the binding free energy for protein-ligand pairs \cite{Osaki_2022}.
In their pyRISM-CNN method, Fowles et al.\ combined a 1D Convolutional Neural Network (CNN) trained on RISM correlation functions with a 1D-RISM solver to predict the solvation free energy and thermodynamic parameters of small organic and drug-like molecules \cite{Fowles_2023}.
\paragraph*{ML-based prediction of RDFs and correlation functions.}
Craven et al.\ used the results of MD simulations to train an ML model based on segmented linear regression and multivariate function decomposition for the prediction of RDFs for pure-component LJ-fluids \cite{Craven_2020}.
Li et al.\ predicted the RDFs of \ce{Ar}, \ce{NO}, and \ce{H2O} from a single MD simulation trajectory using a point cloud-based deep learning strategy \cite{Li_2023}.
Carvalho et al.\ trained a Hopfield neural network (HNN) to solve the OZ equation and to predict RDFs and direct correlation functions for liquid neon, liquid metal and a glassy solid from experimental static structure factors determined by neutron scattering experiments \cite{Carvalho_2020a,Carvalho_2020b,Carvalho_2022a}. Later, these authors also applied a Physics Informed Neural Network (PINN) for the same task to describe argon \cite{Carvalho_2022b}.
To further enhance this PINN approach, Chen et al.\ proposed a PINN tailored for fixed-parameter OZ equations and a physics-informed deep operator network (PIDeepOnet) approach for parameterized OZ equations and reported advances in numerical stability and reduced learning effort \cite{Chen_2024}.
\paragraph*{ML-based prediction of bridge functions.}
Here, the starting point for training data are bridge functions calculated from molecular simulation data using the RDF and cavity correlation function (CCF). Such calculation schemes have been investigated for different model systems, such as HS fluids \cite{Kolafa_2002, Fantoni_2004}, LJ fluids \cite{Kunor_2005, Kunor_2006, Saeger_2016}, Stockmayer fluids \cite{Puibasset_2012}, liquid metals \cite{Lomba_1992, Kambayashi_1994}, or plasma \cite{Poll_1988}.
Bedolla et al.\ used a three-parameter modified Verlet bridge function to describe HS and square-well fluids, where the three parameters were determined by an unconstrained optimization with evolutionary computation algorithms, enforcing the isothermal compressibility to be equal when computed from various routes \cite{Bedolla_2022}.
Goodall et al.\ approximated bridge functions using an ML model to determine pairwise interaction potentials from structural information in terms of correlation functions. This model was used to correct the HNC closure \cite{Goodall_2019, Goodall_2021}. Their feature set comprised the total correlation function, the direct correlation function, the RDF in the form of the fluctuation-dissipation theorem, and the gradient of the indirect correlation function. The training set, including the total correlation function and the static structure factor, was determined by MD simulations employing 13 different interaction potentials. Multi-layer perceptron (MLP) neural networks were trained to approximate the bridge functions.


The goal of this work is to approach exact result using the HNCB equation (\ref{eq:fullHNC}) without relying on one of the above-mentioned approximations for the bridge functions $B(r)$. The bridge functions are learned from exact MC simulations for a fixed interaction $v(r)$ and a range of densities $\rho$ and temperatures $T$. From the MC simulation we obtain the exact RDF $g(r)$ as well as $\gamma(r)$ from the OZ equation (\ref{eq:OZ}) which allows us to invert Eq.~(\ref{eq:fullHNC}) and obtain the exact $B(r)$,
\begin{equation}
    B(r) = \ln g(r) + \beta v(r) - \gamma(r) \ .
    \label{eq:B}
\end{equation}
For small $r$, $v(r)$ becomes very repulsive and accordingly, $g(r)$ becomes very small, which would make MC sampling for small $r$ prohibitively expensive. Therefore, for small $r$ the cavity correlation function (CCF)
\begin{equation}
    y(r)=g(r) \exp[\beta v(r)]
    \label{eq:CCF}
\end{equation}
is calculated {\em directly} (see below) and $B(r)$ can be calculated from
\begin{equation}
    B(r) = \ln y(r) - \gamma(r)
    \label{eq:By}
\end{equation}
Thus the exact RDFs $g(r)$ and the corresponding exact $B(r)$ constitute the learning set for training an artificial neural network (ANN).
We refer to our method, which is based on the exact HNCB closure Eq.~(\ref{eq:fullHNC}) coupled with machine-learned bridge functions, as HNCB-AI.
Remember that the bridge functions $B(r)$ are actually functionals of $g(r)$, i.e.\ $B(r)=B[g;r]$.
The need for such a ``non-local'' dependency of $B(r)$ on the correlations $g$ and $\gamma$ was already stressed in~\cite{Goodall_2019}.

The MC simulations to calculate the RDF $g(r)$, the static structure function $S(k)=1+\rho\int d^3r e^{-i{\bf k}{\bf r}}[g(r)-1]$, and the cavity correlation function $\ln y(r)$ for a set of thermodynamic parameters $\rho$ and $T$ are explained in section~\ref{sec:MC}. We restrict the simulations to the well-studied test bed of statistical physics, the Lennard-Jones liquid, where we give all values in LJ units. The corresponding reduced quantities are indicated with stars, i.e. $T^*=T\,k_\text{B}/\epsilon$ and $\rho^*=\rho\sigma^3$, where $\epsilon$ and $\sigma$ are strength and range of the LJ interaction. The output of the MC simulations is fed in the ANN to train it for $B[g;r]$. The choice of ANN, the learning strategy, and how to solve the coupled HNCB Eq.~(\ref{eq:fullHNC}) and OZ Eq.~(\ref{eq:OZ}) with the $B[g;r]$-ANN are explained in section~\ref{sec:ANN}. Results are discussed in section~\ref{sec:Results}, and conclusions are drawn in section~\ref{sec:Conclusions}.

\section{Molecular Simulations}
\label{sec:MC}

To generate exact correlation functions as the training data for machine learning, we used a MC software suite for molecular simulations \cite{NotebookArchive}. We conducted canonical ensemble (NVT) simulations of $N=\num{2048}$ LJ particles, starting from a face-centered cubic lattice and equilibrating for at least \num{30000} cycles, where each cycle consisted of $N$ attempted single particle moves. Samples were then taken over at least \num{100000} cycles.
We compared the results with simulations of up to \num{20000} molecules and found no significant difference in the correlation functions, which is why we considered $N=\num{2048}$ sufficient.
The maximum translational distance of attempted single particle moves was adjusted to achieve a 50\% acceptance rate.
Statistical uncertainties were estimated with the method proposed by Flyvbjerg \cite{Flyvbjerg_1989}, using at least \num{20} remaining samples.

The RDF was sampled using a histogram method~\cite{allen2017computer}:
%
%
\begin{equation}
    \label{eq:RDFHistogram}
	g( r+\frac{1}{2}\Delta r ) = \left\langle
		\frac{V}{N^2} \frac{1}{V(b)} \sum^{N}_{i=1} \sum^{N}_{j \neq i} h_{\Delta r}(r_{ij})
	\right\rangle_\text{NVT}\ .
\end{equation}
Here, $\Delta r$ is the bin width, $V$ is the total volume, $N$ is the number of sites, $V(b)$ is the volume of a hollow sphere with inner radius $r$ and width $\Delta r$, and $h_{\Delta r}(r_{ij})$ is the characteristic function equal to 1 between $r$ and $r + \Delta r$ and 0 otherwise.
The histogram approximates the RDF at the center $r+\frac{1}{2}\Delta r$ of the interval $[r,r+\Delta r]$.

Additionally, a state of the art force based sampling method for the RDF proposed by Rottenberg \cite{Rotenberg_2020} was used in an attempt to reduce the standard deviation without increasing the simulation length.
It was observed that the results from the force based method had higher standard deviations at radial distances $r > 4\sigma$, which caused inconsistencies during the calculation of the bridge function.
Therefore, only simulation results from histogram sampling were used in this work.

The CCF (\ref{eq:CCF}) can be calculated using the Widom particle insertion method~\cite{Henderson_1983,Llano-Restrepo_1992}:
%
\begin{align}
    \label{eq:CCFSample}
    y(r)
        &= \Big\langle \exp\left[
            - \beta \Psi_{ij}(r)
        \right] \Big\rangle_\text{NVT} 
        \, \Big\langle \exp\left[
            \beta \Psi_{i}
        \right] \Big\rangle_\text{NVT}
        \ .
\end{align}
Here, $\Psi_{i}$ is the total interaction energy of a randomly inserted particle $i$ and $\Psi_{ij}(r)$ is $\Psi_{i}$ without the interaction with a particle $j$ at radial distance $r$.
For small distances $0<r\leq\sigma$, where $g(r)$ is small, we used the directly sampled CCF (\ref{eq:CCFSample}), while we calculate $y(r)$
from $g(r)$ using Eq.~(\ref{eq:CCF}) otherwise.

\begin{figure}
    \centering
    \includegraphics[width=1.\linewidth]{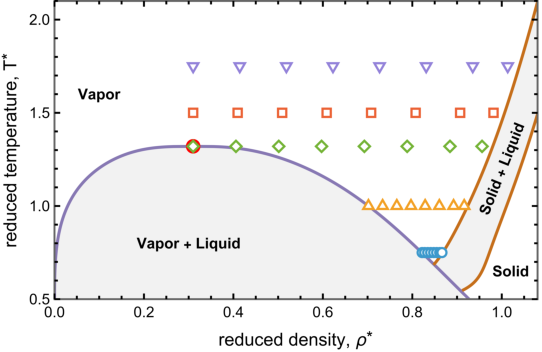}
    \caption{\label{fig:phase_space}
        Points in the phase diagram for which MC calculations were performed. The critical point is marked with a red circle. The phase boundaries, shaded gray, were determined from equations of state \cite{vanderHoef_2000,Thol_2016}.
    }
\end{figure}

The static structure factor $S(k)$ can be sampled directly in $k$-space:
\begin{equation}
    \label{eq:SSF}
    S(k) = \frac{1}{N} \Bigg\langle
        \sum_{n=1}^{N} e^{ i \vec{k} \cdot \vec{r}_n}
        \cdot \sum_{m=1}^{N} e^{ - i \vec{k} \cdot \vec{r}_m}
    \Bigg\rangle_\text{NVT}
\end{equation}
Here, $\vec{r}_n$ is the position of particle $n$, and the wavevector $\vec{k}$ is chosen such that it is consistent with the periodic boundary conditions
$\vec{k} = 2 \pi / L \cdot (n_x,n_y,n_z)$ with integers $n_x$, $n_y$ and $n_z$ and the simulation box side length $L$.
We calculated $S(k)$ directly with (\ref{eq:SSF}) for smaller wavevectors, where the periodic boundary restriction is more important, and
by Fourier transformation of $g(r)$ for larger $k$.

We simulated the LJ fluid for 32 combinations of temperature $T$ and density $\rho$.
Figure~\ref{fig:phase_space} provides an overview of the simulated state points $(T,\rho)$.
With $g(r)$, $y(r)$, and $S(k)$ thus obtained, we calculated the exact bridge function $B(r)$ by first obtaining the indirect correlation functions $\gamma(r)$ from the OZ equation (\ref{eq:OZ}), which is efficiently done by Fourier transformation, and then solved the HNCB equation for $B(r)$ according to Eq.\ (\ref{eq:By}). 

\section{The HNCB-AI method using DeepONets}
\label{sec:ANN}

\subsection{Architecture}

Deep Operator Networks (DeepONets) are neural net architectures designed to learn mappings between function spaces \cite{Lu_2021}. In this study, we employ a DeepONet where the trunk network takes as input a scalar variable $r$, while the branch network processes a function $g(r)$ sampled on a fixed grid with $N_G$ points. The output of the network is a function $B[g; r]$, which depends on both the input variable and the functional input.

The DeepONet consists of two subnetworks: the {\em trunk network} and the {\em branch network}. The trunk network encodes the dependency of the output function on the scalar input $r$. It takes $r$ as input and maps it to a latent representation through a series of fully connected layers with nonlinear activation functions \cite{Lu_2022}. The branch network processes the function $g(r)$ by sampling it on a fixed grid, $\{ r_i \}_{i=1}^{N_G}$, and treating the sampled values $g(r_i)$ as a finite-dimensional input vector. This vector is then passed through multiple layers of a fully connected neural network, extracting relevant features that characterize $g(r)$. For our implementation we used a equidistant grid with $N_G=200$ points up to a distance of $5\sigma$.

The final output of the DeepONet is obtained by taking the inner product of the feature representations from the trunk and branch networks
\begin{equation}
    B[g; r] = \sum_{i=1}^{N_\texttt{net}} \mathcal{T}_i(r) \mathcal{B}_i(g)\ ,
\end{equation}
where $\mathcal{T}_i(r)$ are the outputs of the trunk network and $\mathcal{B}_i(g)$ are the features extracted by the branch network. $N_\texttt{net}=100$ was found sufficient for our application.

This DeepONet formulation is particularly useful in learning solution operators for differential equations where the function $g(r)$ represents input parameters, boundary conditions, or source terms. The model efficiently learns complex functional dependencies, making it a powerful tool for scientific computing and engineering applications \cite{Bhattacharya_2021}.

In this work we employ a slightly different setting, instead of a differential equation we use results of a MC simulation to generate the training data.

\subsection{Training}

Basis for the training were the bridge functions and the radial distribution functions generated with MC for subsets of the densities and temperatures shown in Figure \ref{fig:phase_space}. 
The variable $r$ was also randomly sampled in a range from $0.01\sigma$ to $9\sigma$ with 300 random points uniformly distributed. We added also 200 uniformly distributed points in the region $1\sigma$ to $3\sigma$ since we found that the features of the bridge function here are sensitive for an accurate result of the RDF.

To ensure a correct limiting behavior in the low density limit, we added the zero-density limit $g_\text{0}(r)=f(r)+1$ to the training dataset at temperatures $T^*=$~\numlist{0.5; 0.75; 1.0; 1.25; 1.5; 2.0; 2.5}.

For the training we used an Adam optimizer with a learning rate of \num{0.001} \cite{Kingma_2014}. As loss function we used the mean squared error (squared $L_2$ norm). We trained for \num{60000} epochs and reached a loss below $10^{-4}$.

To demonstrate the predictive power of our approach we used 10 different training sets.  Each training set consists of only 8 randomly selected state points out of the 32 state points obtained with the MC simulations, see Figure \ref{fig:phase_space}. The 10 training sets are given in Table \ref{tab:training_data} in appendix \ref{app:A}. We denote the trained DeepONet with $B_i[g; r]$, where $i$ numbers the training set.
We also trained the DeepONet with all available state point given in Figure \ref{fig:phase_space} and denote the DeepONet with $B_\texttt{LJ}[g; r]$.

\subsection{Solving the HNCB-AI integral equations}
\label{subsec:Solving}


We solve the HNCB-AI equations by generalizing the standard HNC0 implementation, given e.g. by P.B. Warren \cite{SunlightHNC, Warren_2013}:
\begin{enumerate}
    \item Start with a guess for the direct correlation function $c(r)$.
    \item Calculate the indirect correlation function from the OZ relation (\ref{eq:OZ}), which is efficiently done in Fourier space by $\hat\gamma(k)= \frac{\hat c(k)}{1-\rho\hat c(k)} - \hat c(k)$, as described in detail in literature \cite{Llano-Restrepo_1992}.
    \item Fourier transform the indirect correlation function back to real space $\gamma(r)$.
    \item Use the HNCB closure relation to obtain an updated direct correlation function $c_\text{trial}(r)=\exp{[-\beta v(r)+\gamma(r)+B[g;r]]}-\gamma(r)-1$.
    \item To improve convergence, the direct correlation function for the next iteration is obtained by linear mixing of $c_\text{trial}(r)$ with the previous $c(r)$. Continue with 2. until convergence is reached.
\end{enumerate}

The bridge function, needed in step 4, is the prediction with the current RDF calculated as $g(r)=c(r)+\gamma(r)+1$. For efficiency reasons, the $B[g; r]$-net is only updated every 100-th iteration step with the current RDF. To further increase stability of the iteration, we implemented also a linear mixing of the updated bridge function:
\begin{equation}
    B_\text{new}(r)= B_\text{old}(r) +\alpha \left(B[g; r]-B_\text{old}(r)\right) ,
\end{equation}
with $\alpha=0.2$ we reached good convergence. Only for three of the reduced training datasets, i.e.\ 5, 7 and 8 we had a problem with convergence near the critical point. But this seems not to be a problem of the solution method, rather for the combined system of HNC and $B[g; r]$  there is no guarantee of a solution if the DeepONet does not behave well enough. In this work we did not investigate which mathematical conditions are required for a unique solution.
While important, this topic is outside the scope of this work and is left for future exploration.

\section{Results}
\label{sec:Results}

We present results for the RDF $g(r)$ and the pressure $p$ obtained with the HNCB-AI method and compare them with the exact MC results on the one hand and with the simple HNC0 approximation on the other hand.  In section~\ref{ssec:LJ} we stay close to the training set consisting of LJ MC simulations by showing only LJ results, in section~\ref{ssec:HS} we apply the $B[g; r]$-DeepONet for LJ fluids to the HS fluid.

\subsection{Lennard-Jones Liquid}
\label{ssec:LJ}

We incorporate the $B[g; r]$-DeepONet approximation of the bridge function into the HNCB equation~(\ref{eq:fullHNC}) and solve the coupled equations~(\ref{eq:fullHNC}) and~(\ref{eq:OZ}) self-consistently, as explained in section \ref{subsec:Solving}. Since the employed bridge function is only an approximation of the true one, convergence to the correct solution is not guaranteed. Nonetheless, we observe robust convergence across nearly the entire phase diagram, with the exception of regions near the critical point.

As a test of the accuracy of our approach, we applied the HNCB-AI method to the equation of state $p(\rho,T)$. Like essentially all thermodynamic quantities, the pressure $p$ can be calculated from the RDF. We evaluate $p$ using the virial expression
\begin{equation}
\label{eq:Pressure}
p = p_{\text{id}}\left(1 - \frac{2\pi \beta \rho}{3} \int_0^{\infty} \frac{dv}{dr} g(r) r^{3} dr \right)\ ,
\end{equation} 
where $p_{\text{id}} = \rho / \beta$ is the ideal gas pressure. 
Due to its sensitivity to small errors in $g(r)$, Eq.~(\ref{eq:Pressure}) provides a very sensitive test of the predicted RDFs \cite{Reed_1973}.

\begin{figure}
    \centering
    \includegraphics[width=1.1\linewidth]{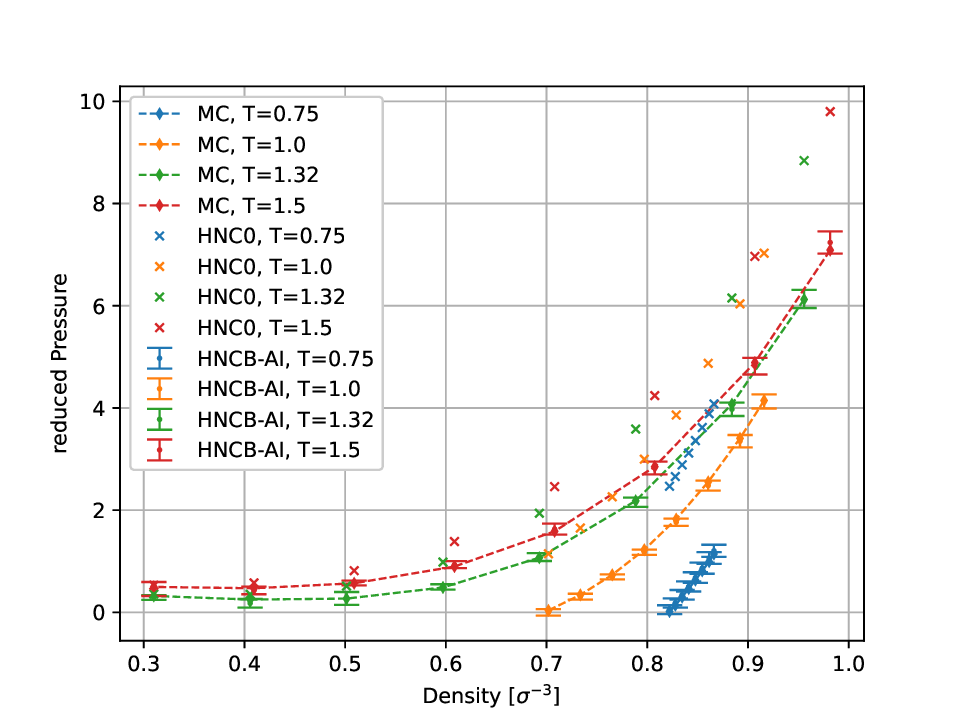}
    \caption{Comparison between the pressure calculated with the DeepONet approximations for the Bridge functionals (symbols with error bars), with the HNC0 approximation (symbols), and with exact MC simulations (lines).}
    \label{fig:pressure_HNC+Br-MC}
\end{figure}

To assess the validity and robustness of our method, we solve the HNCB-AI equations for all 10 models $B_i[g; r]$ and evaluate the pressure and compare with HNC0 and MC results in Figure \ref{fig:pressure_HNC+Br-MC}. The uncertainty can be quantified by the standard deviation of the 10 resulting pressures, indicated by error bars in Figure \ref{fig:pressure_HNC+Br-MC}. The comparison with HNC0 and MC data clearly demonstrates that incorporating the DeepONet-based bridge function leads to a substantial improvement in the predicted pressure. While for all but the lowest densities $\rho$, the HNC0 approximation leads to significantly higher pressures than the exact MC results, the HNCB-AI results are almost indistinguishable from the MC results. The only state point where the uncertainty relative to the pressure value becomes large is near the critical point at $T^*=1.32$ and $\rho^*=0.31$. We investigate this further now.

\begin{figure}
    \centering
    \includegraphics[width=1.1\linewidth]{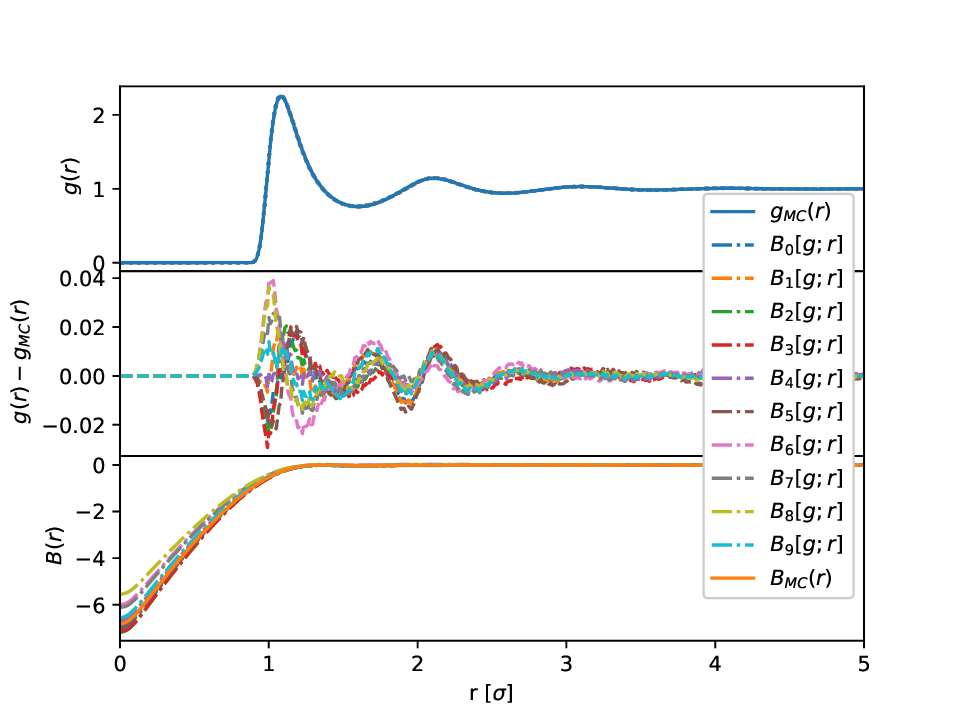}
    \caption{Comparison of the $g(r)$ results from our HNCB-AI method with DeepONet bridge functions (dotted gray lines) and from MC simulations (solid blue lines) for temperature $T=1.32\,\epsilon/k_\text{B}$  and $\rho=0.693\sigma^{-3} $, a density far from the critical one. Only in model 0, the MC data for this density and temperature are in the training set. For all 10 models, the results for $g(r)$ from the HNC+Br method are indistinguishable from the MC result. Therefore we also show the difference multiplied by five (dashed with different colors). The results for the bridge function $B(r)$ are compared to the MC result (dashed-dotted with different colors).}
    \label{fig:gr_HNC+Br-MC_high_density}
\end{figure}

In the top panel of Figure~\ref{fig:gr_HNC+Br-MC_high_density}, we show the RDF $g(r)$ for $T^*=1.32$ and $\rho^*=0.693$, i.e.\ at the critical temperature, but far from the critical density,
cf.\ Figure \ref{fig:phase_space}. The full lines show $g(r)$ (blue) and the associated $B(r)$ (orange) obtained from the MC simulation. The gray dotted lines represent $g(r)$ obtained with our 10 models, and are essentially indistinguishable from the MC result, although only in two of the models (2 and 5), the state point was in the training
set. Because of the small error between our method and the exact result, we also show the difference between the RDFs in the middle panel of Figure~\ref{fig:gr_HNC+Br-MC_high_density}. The deviation of the HNCB-AI $g(r)$ from the exact MC $g(r)$ is less than 2\%.
The DeepONet bridge functions $B(r)$ for the 10 models are shown as dash-dotted lines in the bottom panel. They too match the exact MC $B(r)$ almost perfectly, apart from distances $r$ well below $\sigma$, which are completely irrelevant for the calculation of $g(r)$ because of the repulsive core of the LJ interaction, see Eq.\ (\ref{eq:fullHNC}).

\begin{figure}
    \centering
    \includegraphics[width=1.1\linewidth]{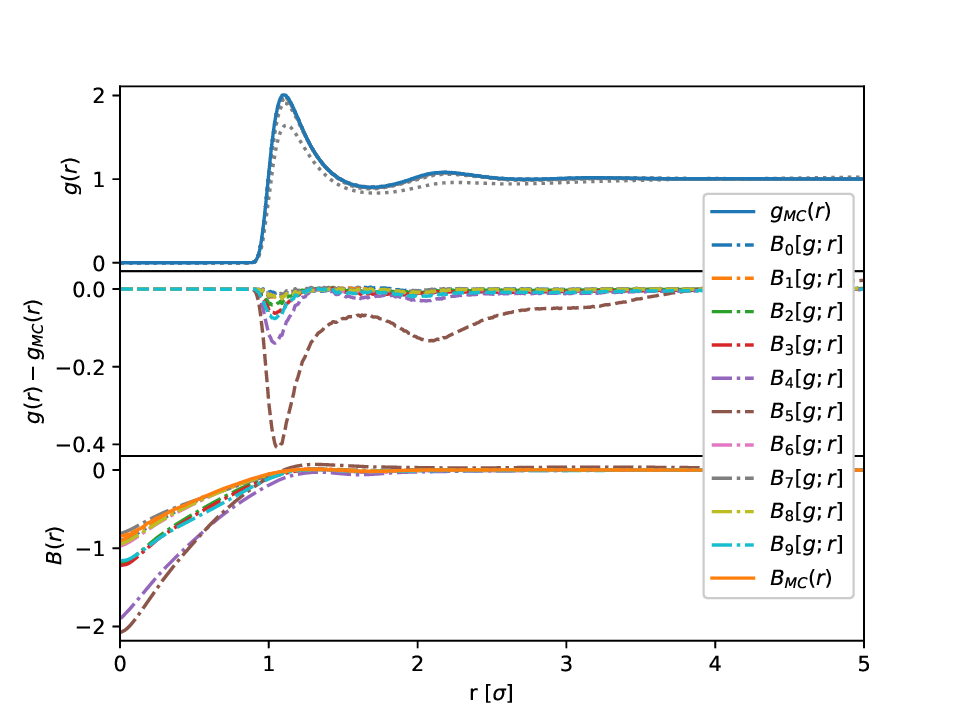}
    \caption{Same as Figure \ref{fig:gr_HNC+Br-MC_high_density}, but for a density close to the critical point, $\rho=0.406\sigma^{-3} $. Only in models 6, 7 and 8 the MC data for this density and temperature are in the training set.}
    \label{fig:gr_HNC+Br-MC_low_density}
\end{figure}

In Figure \ref{fig:gr_HNC+Br-MC_low_density}, we show the same quantities as in Figure \ref{fig:gr_HNC+Br-MC_high_density}, but for
$T^*=1.32$  and $\rho^*=0.406$, which is much closer to the critical density $\rho^*=0.310$, and very close to phase separation, where the homogeneous liquid is unstable, cf.\ Figure \ref{fig:pressure_HNC+Br-MC}. In contrast to the results for the higher density,
not all DeepONets models lead to a stable physical solution for $g(r)$: For models which did not contain this state point in the training set, we see a wide variation of $g(r)$ and $B(r)$ between these models. Only the models which do contain this state point in the training set (models 6, 7, 8) provide reliable results in this case.
This large variation of outcomes results in the large relative uncertainty for the pressure visible in Figure \ref{fig:pressure_HNC+Br-MC}.

This uncertainty is not entirely unexpected. Critical points, where a system undergoes a second-order phase transition, are notoriously hard to calculate quantitatively. Also molecular simulations become cumbersome for a LJ fluid close to the critical point due to large density fluctuations over many length scales. Already our MC input for the DeepONet might be afflicted with finite size effects. A proper treatment would require a finite size scaling analysis~\cite{Binder_2019}, but the detailed study of second order phase transitions is not within the scope of this paper. Instead, we are more interested in using the HNCB-AI method for many different molecules and solvation phenomena, with a wide range of interactions. Therefore, in the following we test HNCB-AI trained with LJ data on a completely different interaction.

\begin{figure}
    \centering
    \includegraphics[width=1.08\linewidth,trim={0 0 0 1cm},clip]{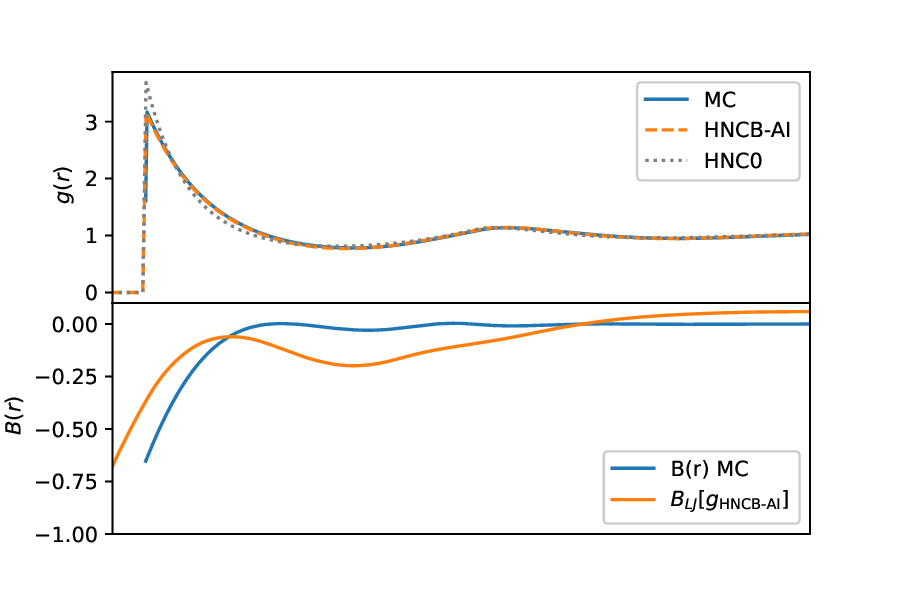}\\
    \vspace{-4ex}
    \includegraphics[width=1.08\linewidth,trim={0 0 0 1cm},clip]{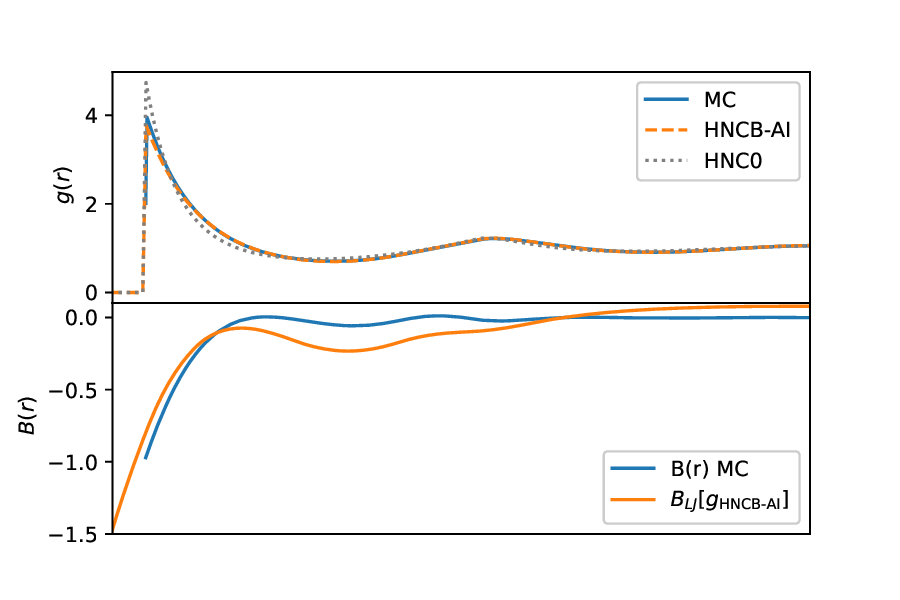}\\
    \vspace{-4ex}
    \includegraphics[width=1.08\linewidth,trim={0 0 0 1cm},clip]{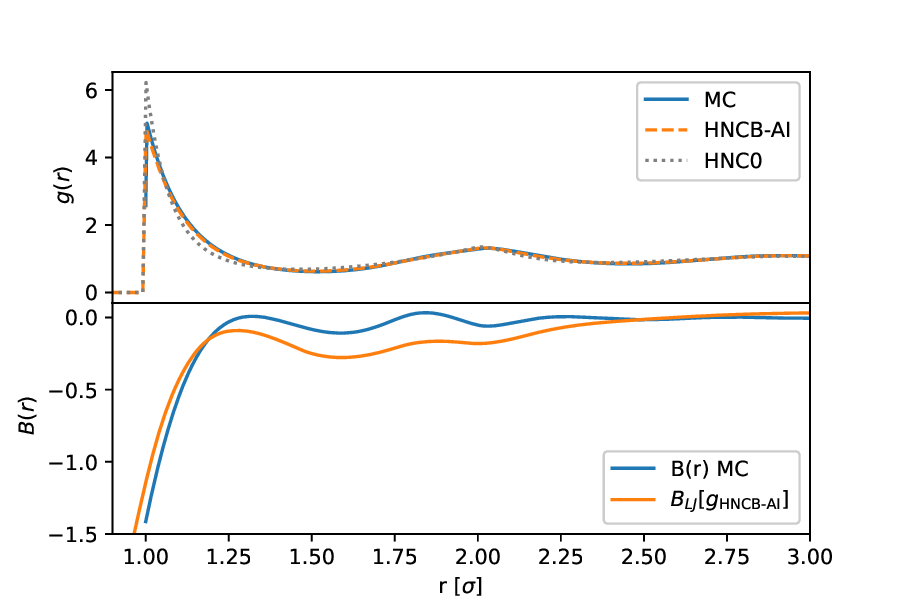}
    \caption{Hard sphere results with $B_\texttt{LJ}[g; r]$ for densities $\rho^*=0.7, 0.8, 0.9$ from top to bottom panel. The top half in each panel compares $g(r)$ obtained with MC simulation (blue), with HNCB-AI using the models trained only on the LJ fluid (orange) and with the HNC0 approximation (grey dashed). The bottom half shows the exact bridge functions $B(r)$ from MC (blue) and from HNCB-AI (orange).
    MC results are taken from Kofala et al.~\cite{Kolafa_2002}.
    }
    \label{fig:gr_HNC+Br-hard_sphere}
\end{figure}

\subsection{Universality of the $B_\texttt{LJ}[g; r]$-DeepONet}
\label{ssec:HS}

In this work, the DeepONet model for the bridge functionals $B_\texttt{LJ}[g; r]$ is based solely on the MC simulation data for the LJ fluid. But since the training data contains no direct information about the interaction potential, one might question the applicability of the functional to other types of interaction. The simplest comparable system is the hard sphere (HS) fluid. We solve the OZ and HNC equations for $g(r)$ for hard spheres with the same interaction range $\sigma$ as in the LJ potential, using $B_\texttt{LJ}[g; r]$. Note that apart from a strong short range repulsion between particles, the physics of the HS fluid is quite different from the LJ fluid: The HS fluid is not a liquid and as such has no liquid-gas phase transition and particularly no critical point; in fact, temperature is irrelevant for the HS fluid.

In Figure \ref{fig:gr_HNC+Br-hard_sphere} we show $g(r)$ for the HS fluid obtained with our HNCB-AI method and compare it with exact MC results \cite{Kolafa_2002} and the HNC0 approximation for the three densities $\rho^*=0.7, 0.8, 0.9$. Our method is in very good agreement with the MC result, except close to the HS radius $\sigma$, where $g(r)$ exhibits a nonanalytic behavior at $r=\sigma$ and where HNCB-AI slightly underestimates the peak. In contrast, the HNC0 approximation deviates from the MC result for the whole range of distances $r$ shown in Figure \ref{fig:gr_HNC+Br-hard_sphere}. This demonstrates that the HNCB-AI method is very robust against changes to a different interaction potential, which is not in the training set and which leads to a qualitatively very different $g(r)$. The explanation for this robustness is the many-to-one mapping from rather different RDFs to very similar bridge functions \cite{Goodall_2019}. This apparent universality of $B[g;r]$ has been long noticed and exploited in approximations where the true bridge functions are replaced by known reference bridge functions \cite{Rosenfeld_1979}. This universality is of course not absolute, but the flexibility of the DeepONet ensures that we find an optimal replacement $B_\texttt{LJ}[g; r]$ for the true $B_\texttt{HS}[g; r]$ without cumbersome adjustment of reference bridge functions.

\begin{figure}
    \centering
    \includegraphics[width=1.01\linewidth]{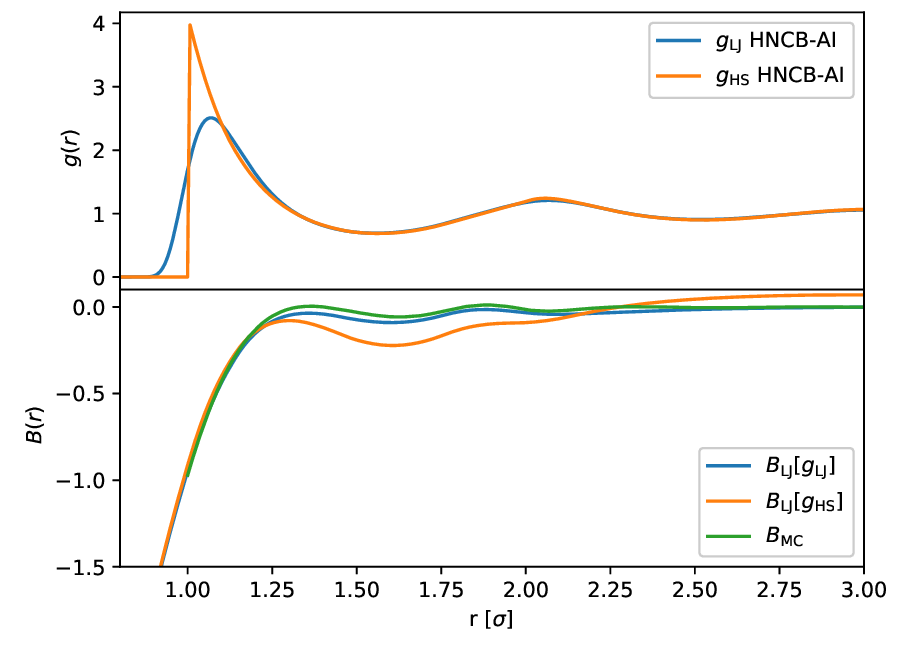}
    \caption{In the upper panel the RDF for the LJ-fluid at $T^*=1.3$ and $\rho^*=0.8$ and the HS-fluid at the same density are compared, which are input to the $B_\texttt{LJ}[g; r]$ net. The result is compared to the MC HS bridge function in the lower panel.}
    \label{fig:Br_LJ_vs_HS}
\end{figure}

Indeed, compared to $g(r)$, $B(r)$ is a fairly featureless function and looks rather similar for a LJ liquid and a HS fluid. In Figure \ref{fig:Br_LJ_vs_HS} we compare the output of our $B_\text{LJ}[g; r]$ DeepONet for a HS and LJ input. We observe that indeed the prediction for the bridge function with the LJ input, $B_\text{LJ}[g_\text{LJ}; r]$ is very close to the MC result. If we use the HS input, the agreement below $1.25\sigma$ is still excellent, but for larger values a substantial difference is visible, which is not surprising as the training data does not contain any information on HS fluids. What is surprising, however, is that the difference in this region has only little influence on the resulting $g(r)$, since part of it is canceled by the indirect correlation function $\gamma(r)$ as we show in Figure \ref{fig:deconstruction_RDF} in appendix \ref{app:B}. 

\section{Conclusions}
\label{sec:Conclusions}

This study investigates whether it is possible to train an artificial neural network to predict the bridge functions, which is the missing information in the hypernetted-chain (HNC) closure of the integral equation theory to describe a classical fluid exactly. We generate learning data in the form of the RDF $g(r)$ for a single component Lennard-Jones fluid, with MC simulations which we use as input for a deep operator network, DeepONet. The learning target is the bridge functional $B[g;r]$ which is obtained from $g(r)$ by solving the Ornstein-Zernike (OZ) equation~(\ref{eq:OZ}) and the HNCB equation (\ref{eq:B}). The key idea of our method, HNCB-AI, is that the bridge functional has much less structure than other correlation functions and constitutes only a correction to the HNC0 approximation where $B[g;r]$ is neglected. Our results clearly demonstrate the capability of this approach, not only within one type of interaction, but also for related interactions {\em without} additional training.

We simulated the LJ fluid with MC for a range of temperatures and densities $(T,\rho)$, including the critical point, from which we obtain exact RDFs $g(r)$, the related static structure functions $S(k)$, and also the cavity correlation function, required for an accurate calculation of $B[g;r]$. From the exact results we randomly drew 10 training sets, each consisting of a quarter of all simulated state points $(T,\rho)$. We trained a DeepONet which provides 10 models for $B[g;r]$ which we used to self-consistently solve the OZ equation (\ref{eq:OZ}) and the HNCB equation (\ref{eq:fullHNC}) to calculate $g(r)$. The quality of the models was tested by comparing the pressure calculated by the virial expression with the exact MC results for all state points. We find very good agreement, except near the critical point, where we observe a larger variance between the models. Near the critial point, we are close to the spinodal instability where a small change in $T$ of $p$ would break up the liquid. Small deviations thus have a big effect in this region, and the OZ and HNCB equations are harder to solve. Also the MC results are less reliable there due to finite size effects which become very important at the critial point. The result is of course considerably improved if the model contains the critical point in the training data.

For a more detailed analysis, we compare the HNCB-AI and MC results for $g(r)$ and $B[g;r]$, which confirms the excellent agreement except near the critical point where $g(r)$ deviated significantly in one of the models which did not include the critical point. We conjecture that if the training set contains the ``problematic'' state points, like critical points of a LJ fluid, our HNCB-AI method offers a quantitative description of fluids that rivals much more time-consuming exact MC simulations.

We examined the robustness of the HNCB-AI method by employing the models based only on the LJ fluid to a different fluid, composed of hard spheres with the same interaction range $\sigma$ as the LJ particles. Even for very high density, the DeepONet bridge functional learned from the LJ simulations proved to be very universal. The RDFs $g(r)$ obtained with these bridge functionals are almost indistinguishable from the exact $g(r)$, except close to $r=\sigma$ where $g(r)$ has a discontinuity.


Our promising results warrant further investigations into whether HNCB-AI could be a general purpose closure for integral equation theories, usable for example for the RISM method. The robustness of HNCB-AI will be tested with a larger variety of interactions, including Coulomb interactions crucial for liquids of molecules with partial charges. Mixtures of solvents and solutes are of interest for chemical applications, and for that we need to assess the applicability of our scheme to complex molecules and intermolecular interactions with different length scales, unlike the single length scale $\sigma$ present in the single-component LJ or HS fluid. Furthermore, it will be beneficial to incorporate ,,blind'' quality checks, such as thermodynamic self-consistency \cite{pellicaneFluidPhaseEquil20}, to gauge the validity of HNCB-AI results without doing expensive MC simulations every time a new molecule is studied which is not contained in the learning data.

\begin{acknowledgments}
Calculations were partially performed using supercomputer resources provided by the Vienna Scientific Cluster (VSC), project no.\ 72457. The authors gratefully acknowledge support from NAWI Graz.
\end{acknowledgments}

\appendix

\section{Subsets of training data}
\label{app:A}

Our training set consists of 32 MC simulations for a range of temperatures $T$ and densities $\rho$, including the critical point of the liquid-gas phase transition, see Figure \ref{fig:phase_space}. We assess the quality of our HNCB-AI method by using 10 different training data sets, each consisting of 8 randomly chosen phase points $(T,\rho)$, i.e.\ 25\% of the full set of MC simulations, resulting in 10 DeepONet models $B_i[g; r]$, where $i=0,\dots,9$ denotes the model. This allows us to collect statistics and provide the error bounds in Figure \ref{fig:pressure_HNC+Br-MC} for the equation of state $p(\rho)$ obtained with HNCB-AI. 
In Table \ref{tab:training_data} we list the randomly chosen training data sets for the 10 models.

\begin{table}
    \centering
    \caption{\label{tab:training_data}List of randomly selected MC data for training the 10 DeepONet models $B_i[g; r]$. }
    \begin{tabular}{|rr|l|l|l|l|l|l|l|l|l|l|}
    \hline
    Temp & Density & 0 & 1 & 2 & 3 & 4 & 5 & 6 & 7 & 8 & 9 \\
    \hline\hline
    0.750 & 0.822 &  &  &  &  &  & x &  &  & x &  \\
    0.750 & 0.828 &  &  & x & x & x &  &  &  &  &  \\
    0.750 & 0.835 &  &  & x & x & x &  &  &  &  &  \\
    0.750 & 0.841 & x &  &  &  & x &  & x &  &  &  \\
    0.750 & 0.848 &  &  &  &  &  &  &  &  &  & x \\
    0.750 & 0.855 &  &  &  &  &  & x &  & x & x &  \\
    0.750 & 0.861 & x &  &  &  &  &  &  & x &  &  \\
    0.750 & 0.866 &  &  &  &  &  &  & x & x &  &  \\
    1.000 & 0.702 & x &  &  &  &  &  &  &  &  &  \\
    1.000 & 0.734 &  &  & x & x &  &  &  & x &  & x \\
    1.000 & 0.765 &  &  &  &  &  & x & x &  &  &  \\
    1.000 & 0.797 &  &  &  &  &  &  &  &  &  &  \\
    1.000 & 0.829 &  &  &  &  & x & x &  &  & x &  \\
    1.000 & 0.860 &  & x &  &  &  &  &  &  &  &  \\
    1.000 & 0.892 &  &  &  &  &  &  &  &  &  &  \\
    1.000 & 0.916 &  &  &  &  &  &  & x &  &  &  \\
    1.320 & 0.310 &  &  & x & x & x &  &  &  &  &  \\
    1.320 & 0.406 &  &  &  &  &  &  & x & x & x &  \\
    1.320 & 0.501 & x & x &  &  &  & x & x & x &  &  \\
    1.320 & 0.597 &  &  & x & x & x &  &  &  &  &  \\
    1.320 & 0.693 & x &  &  &  &  &  &  &  &  &  \\
    1.320 & 0.789 &  &  &  &  &  & x &  &  &  &  \\
    1.320 & 0.884 &  &  &  &  &  &  & x &  &  &  \\
    1.320 & 0.956 &  & x &  &  & x &  &  &  &  & x \\
    1.500 & 0.310 & x &  &  &  &  & x & x &  & x & x \\
    1.500 & 0.409 &  & x &  &  &  &  &  &  &  & x \\
    1.500 & 0.509 & x & x &  &  &  &  &  &  & x & x \\
    1.500 & 0.609 &  &  &  &  &  & x &  &  &  &  \\
    1.500 & 0.708 & x &  & x & x &  &  &  &  &  &  \\
    1.500 & 0.807 &  & x &  &  &  &  &  & x &  &  \\
    1.500 & 0.907 &  & x & x & x &  &  &  &  & x & x \\
    1.500 & 0.982 &  & x & x & x & x &  &  & x & x & x \\
    \hline
    \end{tabular}
\end{table}

\section{Sensitivity of the RDF on the bridge function}
\label{app:B}

\begin{figure}
    \centering
    \includegraphics[width=1.1\linewidth]{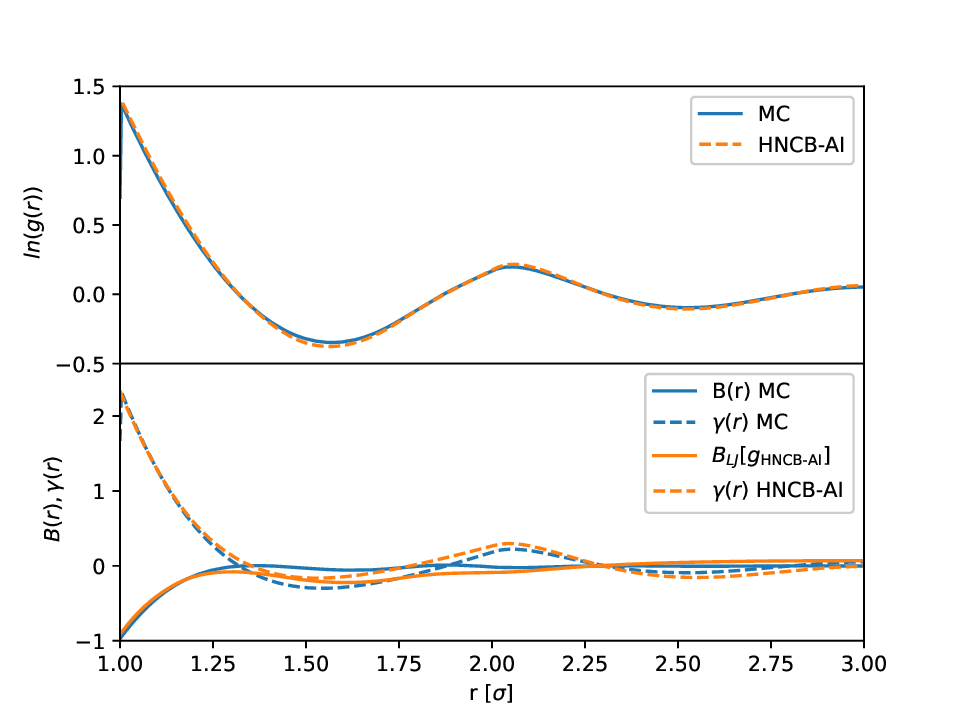}
    \caption{In the upper panel, the logarithm of the RDF is shown for MC and HNCB-AI for $\rho^*=0.8$, which is the sum of the bridge function and the indirect correlation function $\gamma(r)$, plotted in the lower panel. It can be seen that the differences in the bridge functions are largely canceled by the corresponding indirect correlation function.}
    \label{fig:deconstruction_RDF}
\end{figure}

As can be seen in Figure \ref{fig:gr_HNC+Br-hard_sphere}, there are remarkable differences in the hard sphere bridge function between the MC result \cite{Kolafa_2002} and the prediction of the $B_\text{LJ}$ net, whereas the resulting RDFs show very good agreement. In Figure~\ref{fig:deconstruction_RDF} we separate the contributions to the RDF $g(r)$ into the bridge function and the indirect correlation $\gamma(r)$, see Eq.\ (\ref{eq:fullHNC}), for $\rho^*=0.8$. We see that the differences in the bridge functions are largely canceled by opposite differences in $\gamma(r)$, such that the resulting RDF $g(r)$ agrees much better with the exact MC result than the individual contributions.

\bibliography{citations}

\end{document}